\documentclass[12pt]{iopart}

\usepackage{graphicx}
\usepackage{color}
\begin{document}

\title{Topological phase transition induced by magnetic proximity effect in two dimensions}
\author{Yijie Zeng$^1$, Luyang Wang$^{1,*}$, Song Li$^2$, Chunshan He$^1$, Dingyong Zhong$^1$ and Dao-Xin Yao$^{1,**}$}

\address{$^1$ State Key Laboratory of Optoelectronic Materials and Technologies, School of Physics, Sun Yat-sen University, Guangzhou 510275, China}
\address{$^2$ Department of Mechanical and Biomedical Engineering, City Unversity of Hong Kong, Kowloon, China}
\eads{$^*$\mailto{wangluyang730@gmail.com}\\$^{**}$\mailto{yaodaox@mail.sysu.edu.cn}}
\vspace{10pt}
\begin{indented}
\item[]April 2019
\end{indented}

\begin{abstract}
We study the magnetic proximity effect on a two-dimensional topological insulator in a CrI$_3$/SnI$_3$/CrI$_3$ trilayer structure. From first-principles calculations, the BiI$_3$-type SnI$_3$ monolayer without spin-orbit coupling has Dirac cones at the corners of the hexagonal Brillouin zone. With spin-orbit coupling turned on, it becomes a topological insulator, as revealed by a non-vanishing $Z_2$ invariant and an effective model from symmetry considerations. Without spin-orbit coupling, the Dirac points are protected if the CrI$_3$ layers are stacked ferromagnetically, and are gapped if the CrI$_3$ layers are stacked antiferromagnetically, which can be explained by the irreducible representations of the magnetic space groups $C_{3i}^1$ and $C_{3i}^1(C_3^1)$, corresponding to ferromagnetic and antiferromagnetic stacking, respectively. By analyzing the effective model including the perturbations, we find that the competition between the magnetic proximity effect and spin-orbit coupling leads to a topological phase transition between a trivial insulator and a topological insulator.
\end{abstract}

%
\vspace{2pc}

\section{Introduction}
Topological insulators (TIs) are time-reversal invariant systems with a non-zero $Z_2$ index\cite{PhysRevLett.95.146802,RevModPhys.82.3045}. They are extraordinary in that a gapless, conducting surface (edge) state exists while the bulk is insulating. The necessary condition for TIs to occur is band inversion induced by spin-orbit coupling (SOC). The first model for two-dimensional (2D) TIs is graphene\cite{PhysRevLett.95.226801}, and later confirmed in HgTe/CdTe heterostructure\cite{Bernevig1757,doi:10.7566/JPSJ.82.102001}.
To detect the edge state a sizeable bulk band gap is necessary. In the search of new 2D TIs candidate materials, two kinds of efforts are made, one is to increase the SOC in graphene, by forming heterostructure with materials with large SOC\cite{doi:10.1021/acsnano.6b05982,graphene1,g2}. The other is to predict other TIs with large gap by first-principles calculation, such as germanene\cite{PhysRevLett.107.076802}, tin film\cite{PhysRevLett.111.136804}, bismuth bilayer\cite{PhysRevB.83.121310} and distorted $1T^{'}$ transition-metal dichalcogenide MX$_2$ (M$=$ Mo, W and X$=$ S, Se)\cite{Qian1344}.

In 2D TIs, the edge state causes quantum spin Hall effect (QSHE), which is the analogue of quantum Hall effect. When there is spontaneous magnetization, the exchange interaction between the conduction electron in the edge state and the magnetic moment will gap the edge state\cite{magneticTI}, leading to quantum anomalous Hall effect (QAHE)\cite{Yu61,PhysRevLett.106.166802}. The spontaneous magnetization can be introduced by doping magnetic transition metals into the TIs\cite{Chen659,Lee1316}. However, the random distribution of dopant can greatly affects the gap. Another way is to use the magnetic proximity effect by forming heterostructure with ferromagnetic insulators\cite{FM_TI1,FM_TI2} while a well-matched heterostructure is crucial.

Here we concentrate on the second way of introducing magnetism into a TI, by constructing a model CrI$_3$/SnI$_3$/CrI$_3$ trilayer heterostructure. CrI$_3$ is a layered ferromagnet with Curie temperature of $61$ K\cite{doi:10.1021/cm504242t}. In the monolayer it is Ising ferromagnet with magnetic moment perpendicular to the layer, while bilayer CrI$_3$ shows antiferromagnetic ground state\cite{CrI3_Monolayer,CrI3_C}. By forming heterostructure with SnI$_3$, which is shown to be a TI, it is found that ferromagnetic alignment between CrI$_3$ layers preserves the Dirac point of SnI$_3$, while antiferromagnetic alignment opens a gap of about $16$ meV. Furthermore, ferromagnetic alignment introduces weak Van Vleck paramagnetism in SnI$_3$, and turning it into a magnetic TI. An effective Hamiltonian including both SOC and antiferromagnetism is established, by including spin into the Haldane model\cite{Haldane}. In such a system phase transition between a trivial antiferromagnetic insulator and TI can occur, depending on the relative strengths of SOC and antiferromagnetism.  Our findings may provide some insight to the spin-orbit torque effect\cite{EDELSTEIN1990233} at the ferromagnet/TI interface\cite{SOT1,SOT2,PhysRevLett.119.077702}, and the resulting current-induced magnetization switching by magnetic proximity effect. The electronic structure of SnI$_3$ monolayer is discussed in Section \ref{sec:SnI3}, followed by the magnetic proximity effect in CrI$_3$/SnI$_3$/CrI$_3$ trilayer structure. A brief conclusion is given in Section \ref{conclusion}.

\section{Structure and electronic properties of SnI$_3$ monolayer}
\label{sec:SnI3}
\subsection{Results from first-principles calculations}
\begin{figure}
	\includegraphics[width=8cm]{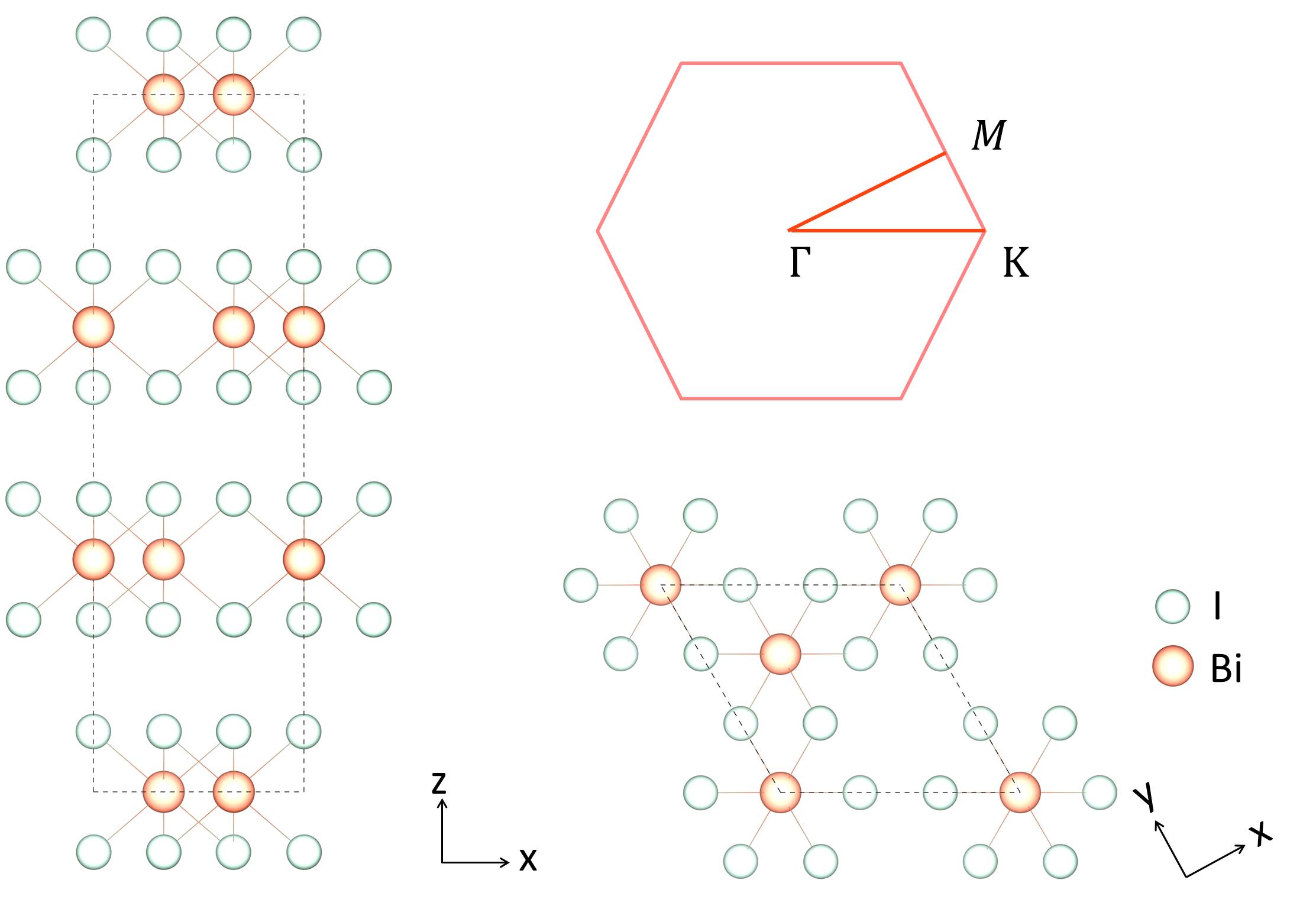}
	\centering
	\caption{The unit cell of BiI$_3$ structure seen on $(1\overline{2}0)$ (Left), $(001)$ (Right bottom) surface, and the Brillouin zone for the monolayer form (Right up).}
	\label{fig:struc}
\end{figure}

The bulk SnI$_3$ is in rhombohedral BiI$_3$ structure\cite{doi:10.1063/1.4813486,doi:10.1063/1.4932129,NASON1995221} (space group $R\overline{3}$, $C_{3i}^2$, No.$148$), the unit cell of which can be treated in two ways. One is using rhombohedral axes, where the three basis vectors have the same length and the same angles between each other. The other is using hexagonal axes, where the basis vectors $\vec{a}$, $\vec{c}$ are the same as those of hexagonal crystals. Here we use the second since it is convenient to reveal the layer structure of BiI$_3$. As a result, each unit cell contains three layers of BiI$_3$, connected by weak van de Waals interaction, and each layer of BiI$_3$ is composed of three atomic layers -- two iodine layers and one tin layer in between, see Figure \ref{fig:struc}. There is a displacement of $(\vec{a}+\vec{b})/3$ between neighboring layers, where $\vec{a}$ and $\vec{b}$ are basis vectors of the unit cell.
\begin{figure}
    \includegraphics[width=8cm]{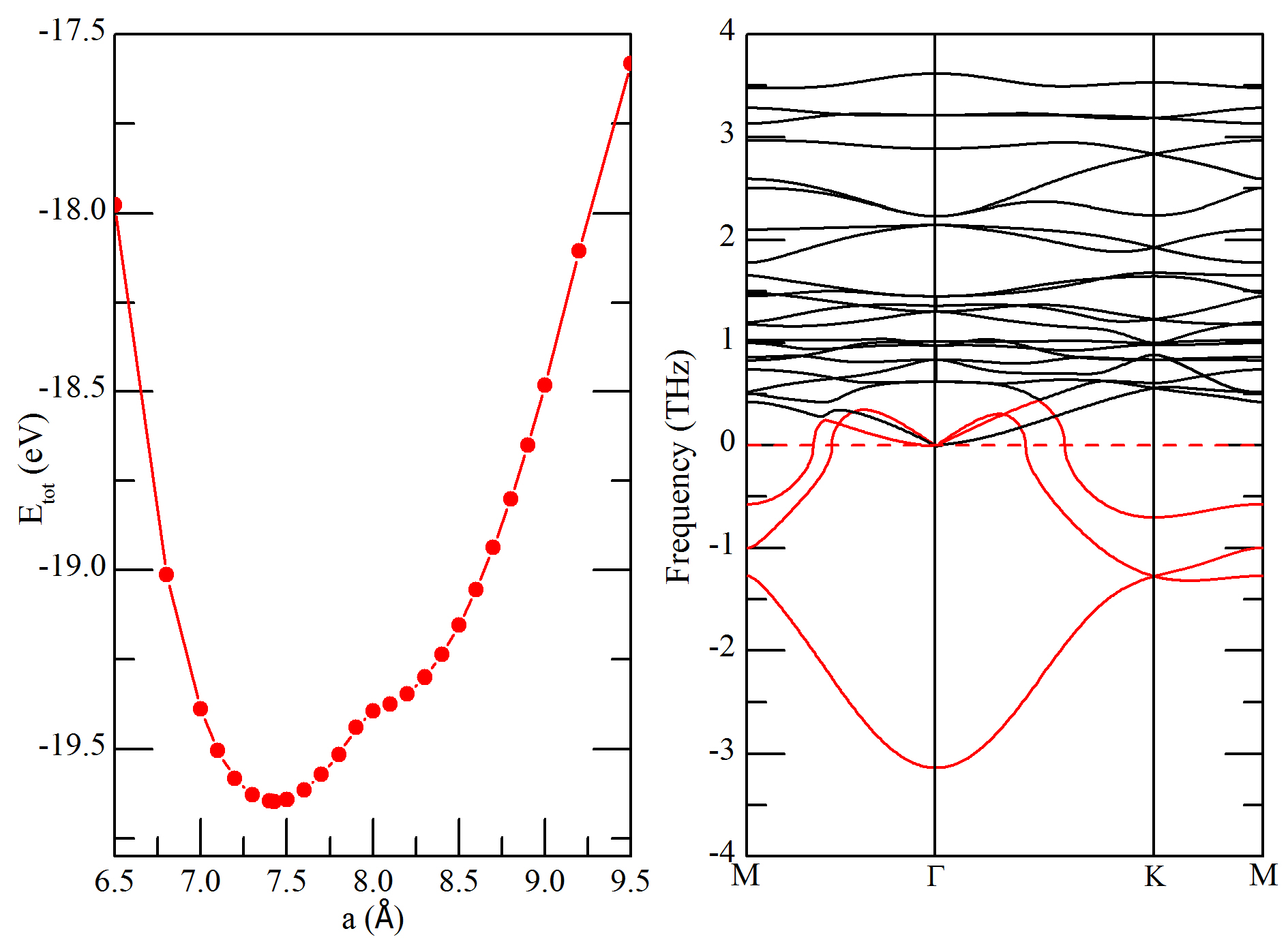}
    \centering
    \caption{(Left) The total energy without SOC versus lattice constant for monolayer SnI$_3$. (Right) Phonon dispersion of monolayer SnI$_3$, the red curves correspond to imaginary modes.}
    \label{fig:E-a}
\end{figure}

Compared with the bulk, which has the $C_{3i}$ point group symmetry, the monolayer has an additional twofold rotation symmetry perpendicular to $\vec{c}$, and hence belongs to the $D_{3d}$ point group. In the early stage of our analysis, the SnI$_3$ monolayer was supposed to be grown on Ag$(111)$ surface, leading to a lattice constant of $8.7$ \AA.

\begin{figure}[ht]
	\includegraphics[width=8cm]{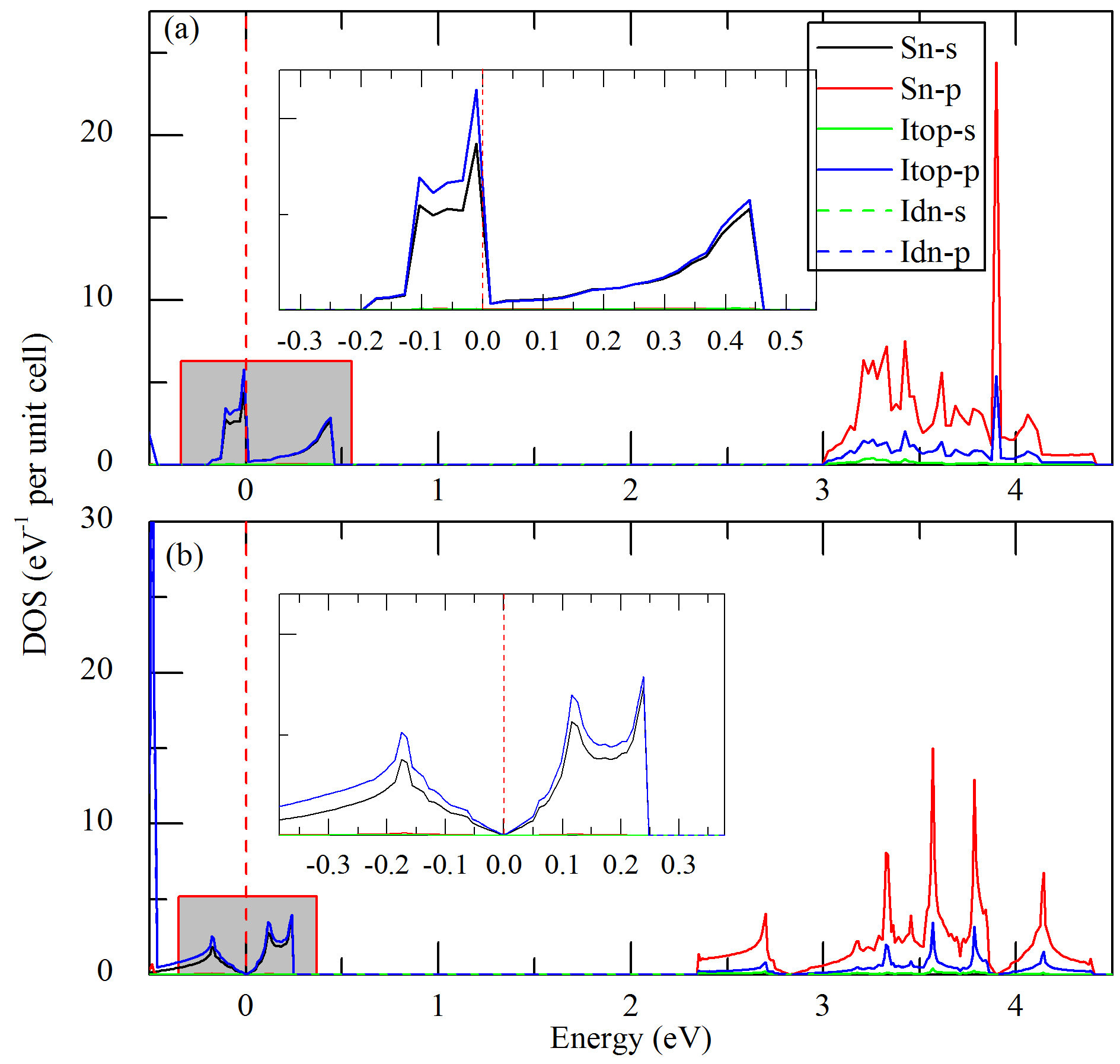}
	\centering
	\caption{Density of states for SnI$_3$ monolayer at lattice constant of (a) $7.0$ \AA \ and (b) $8.7$ \AA. The insets are the enlarged region around the Fermi energy.}
	\label{fig:DOS}
\end{figure}

To see whether strain exists at this lattice constant, we have calculated the total energies under different lattice constants, as shown in Figure \ref{fig:E-a}. The lattice constant with lowest energy is $7.43$ \AA. Thus a tensile strain about $17\%$ would exist in this monolayer if it were grown on Ag(111) surface. Furthermore, the phonon dispersion calculated at $8.7$\AA \ shows an imaginary optical and two imaginary acoustic phonon branches (Figure \ref{fig:E-a}), indicating the monolayer might be unstable at low temperature. Hereafter we will not discuss the stability issue but pay attention to its electronic properties.
\begin{figure*}[ht]
	\includegraphics[width=16cm]{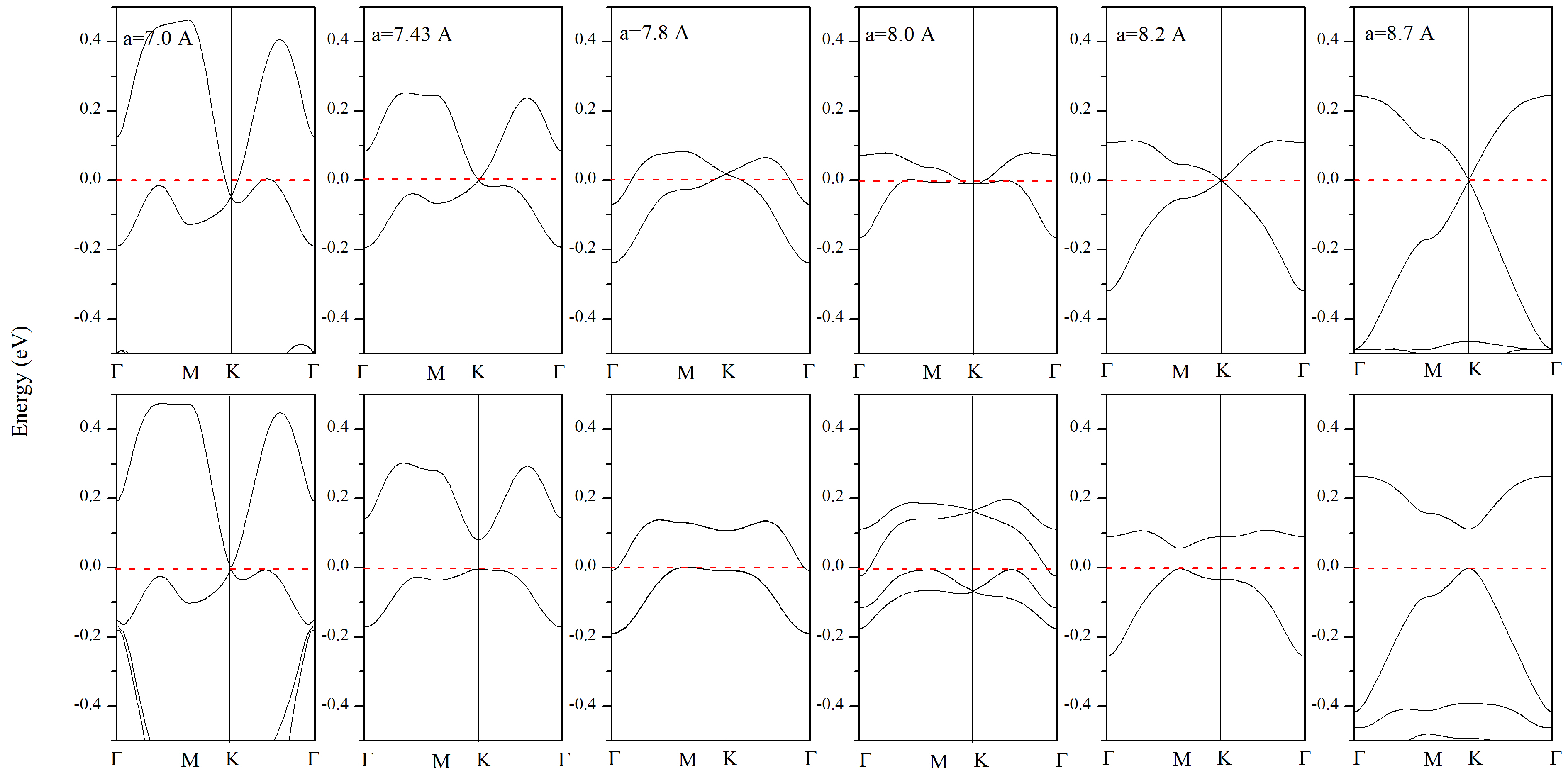}
	\centering
	\caption{The band structure of the SnI$_3$ monolayer without (Up) and with (Bottom) SOC, at different lattice constants. Only the two bands around the Fermi energy are shown.}
	\label{fig:band_structure}
\end{figure*}
The band structures without and with SOC of the SnI$_3$ monolayer at selected lattice constants are shown in Figure \ref{fig:band_structure}. Two bands, which come from the I $p$ and Sn $s$ states, appear near the Fermi energy ($E_F$), as shown in Figure \ref{fig:DOS}. Without SOC, the two bands touch at $K$, realizing a two dimensional representation of the group of the wave vector at $K$ ($C_{3v}$). Especially, at lattice constant of $7.43$ \AA, $8.2$ \AA\ and $8.7$ \AA, the energy dispersion near $K$ is linear, and $E_F$ locates exactly at the touching point, leading to a vanishing density of state (DOS). After including the SOC effect, which is supposed to be large due to the presence of the heavy iodine\cite{Slater}, a gap of $0.11$ eV opens in the Brillouin zone. Due to time-reversal symmetry and inversion symmetry, the bands are doubly degenerate at any $k$ even with SOC\cite{PhysRev.52.361}. (There is an anomaly at $8.0$ \AA, in which case magnetism appears after including SOC, and the double degeneracy is broken. This spontaneous magnetization causes the inflection in the $E-a$ curve near $8.0$ \AA, as shown in Figure \ref{fig:E-a} (a).) At other lattice constants, there is a finite DOS at $E_F$ and the system is a compensated semimetal.

\subsection{Effective model}
The gaplessness without SOC and the SOC induced gap also appear in graphene\cite{RevModPhys.81.109,PhysRevB.75.041401,PhysRevB.74.165310}, which is a semimetal without SOC and a 2D TI once the intrinsic SOC is taken into account. This leads us to ask whether the SnI$_3$ monolayer is a 2D TI. The calculated $Z_2$ invariant is $1$ (see appendix), indicating the system is a TI. Here we develop the effective theory near the degenerate point. The generic Hamiltonian near $K$ to the first order of $\vec{k}$ can be written as
\begin{equation}
	H_{\vec{K}}(\vec{k})=c_ik_i+\vec{R}(\vec{k})\cdot\vec{\sigma}
\end{equation}
where $\vec{R}(\vec{k})$ is a vector with each component linear in $k_i$, and $\sigma_i$'s are the Pauli matrices of a pseudospin associated with the two bands near the Dirac point $K$. The energy spectrum is a Dirac cone. Due to the $C_{3v}$ symmetry, the Dirac cone must be isotropic, so we can write $H_{\vec{K}}(\vec{k})=v_F\vec{k}\cdot\vec{\sigma}$. Here $v_F$ is the Fermi velocity, which is about $1.47\times 10^5$ m/s at lattice constant of $8.7$\AA, deduced from the band structure. The magnitude of Fermi velocity is of interest for creation of bound states\cite{Fermi_vel}. Time reversal requires that $H_{-\vec{K}}(\vec{k})=H_{\vec{K}}^*(-\vec{k})=-v_F\vec{k}\cdot\vec{\sigma}^*$. The full Hamiltonian that describes the low energy states near both $\vec{K}$ and $-\vec{K}$ is
\begin{equation}
H(\vec{k})=
\left( \begin{array}{cc}
H_{\vec{K}}(\vec{k}) & 0 \\
0 & H_{-\vec{K}}(\vec{k}) \\
\end{array}
\right)
=v_F(k_x\sigma_x\tau_z+k_y\sigma_y)
\end{equation}
which satisfies the time reversal symmetry $TH(\vec{k})T^{-1}=H(-\vec{k})$. Here $T=i\tau_xs_yK$ is the time reversal operation including spin, with $\tau_i$'s being the Pauli matrices for the valley degree of freedom and $s_i$'s the Pauli matrices for spin. Since the system also has inversion symmetry, that is, $PH(\vec{k})P=H(-\vec{k})$, we can deduce the matrix for the inversion operator $P=\sigma_x\tau_x$ with awareness of that inversion exchanges $\vec{K}$ and $-\vec{K}$, and does not affect the spin. The SOC term compatible with both time reversal and inversion symmetry, as well as the roational symmetry of the lattice is $H_{SO}=\lambda\sigma_z\tau_zs_z$\cite{PhysRevLett.95.226801}. The Rashba SOC is prohibited by the inversion symmetry. The chiral symmetry is broken with the addition of the SOC term, which is intrinsically linked to topological phases. The Hamiltonian is the same as the effective Hamiltonian of Kane-Mele model\cite{PhysRevLett.95.146802} without the Rashba term, so we expect the system is a $Z_2$ TI in 2D.

\section{Magnetic proximity effect in CrI$_3$/SnI$_3$/CrI$_3$ trilayer structure}
\label{sec:trilayer}
\subsection{Results from first-principles calculations}
Having confirmed the SnI$_3$ monolayer is a TI, we then construct a CrI$_3$/SnI$_3$/CrI$_3$ trilayer structure, in the same stacking sequence as in bulk BiI$_3$, to study the magnetic proximity effect on the TI. This newly designed heterostructure belongs to space group $P\overline{3}$ ($C_{3i}^1$, No.$147$). The lattice constant is chosen to be $7.43$ \AA, under which the CrI$_3$ monolayer suffers a tensile strain about $8\%$. However, the ferromagnetic ordering is still the ground state under this tensile strain\cite{PhysRevB.98.144411}. Therefore, the Cr magnetic moments in each CrI$_3$ layer are aligned ferromagnetically, perpendicular to the layer, while both ferromagnetic and antiferromagnetic alignment between the CrI$_3$ layers are considered. Several experimental works on the monolayer-substrate heterostructure have shown large lattice mismatch might exist for 2D materials. CrI$_3$/WSe$_2$ heterostructure has been fabricated to study the valley splitting of WSe$_2$ due to magnetical exchange field of CrI$_3$\cite{strain}, where the lattice mismatch is $5.4\%$\cite{MoSe2_CrI3}. Another example is monolayer FeSe deposited on SrTiO$_3$(110) plane, where one of in-plane lattice constants is reduced by $6\%$\cite{FeSe}. Earlier work on MoS$_2$ even demonstrates strain as large as $11\%$\cite{MoS2B}. Moreover, the Dirac point of monolayer SnI$_3$ exists in the whole range of lattice constants we studied, as a result of the $D_{3d}$ symmetry. Thus we expect the lattice constant chosen for the heterostructure is reasonable and would not affect the main conclusion presented in the following.

\begin{figure*}[ht]
	\includegraphics[width=16cm]{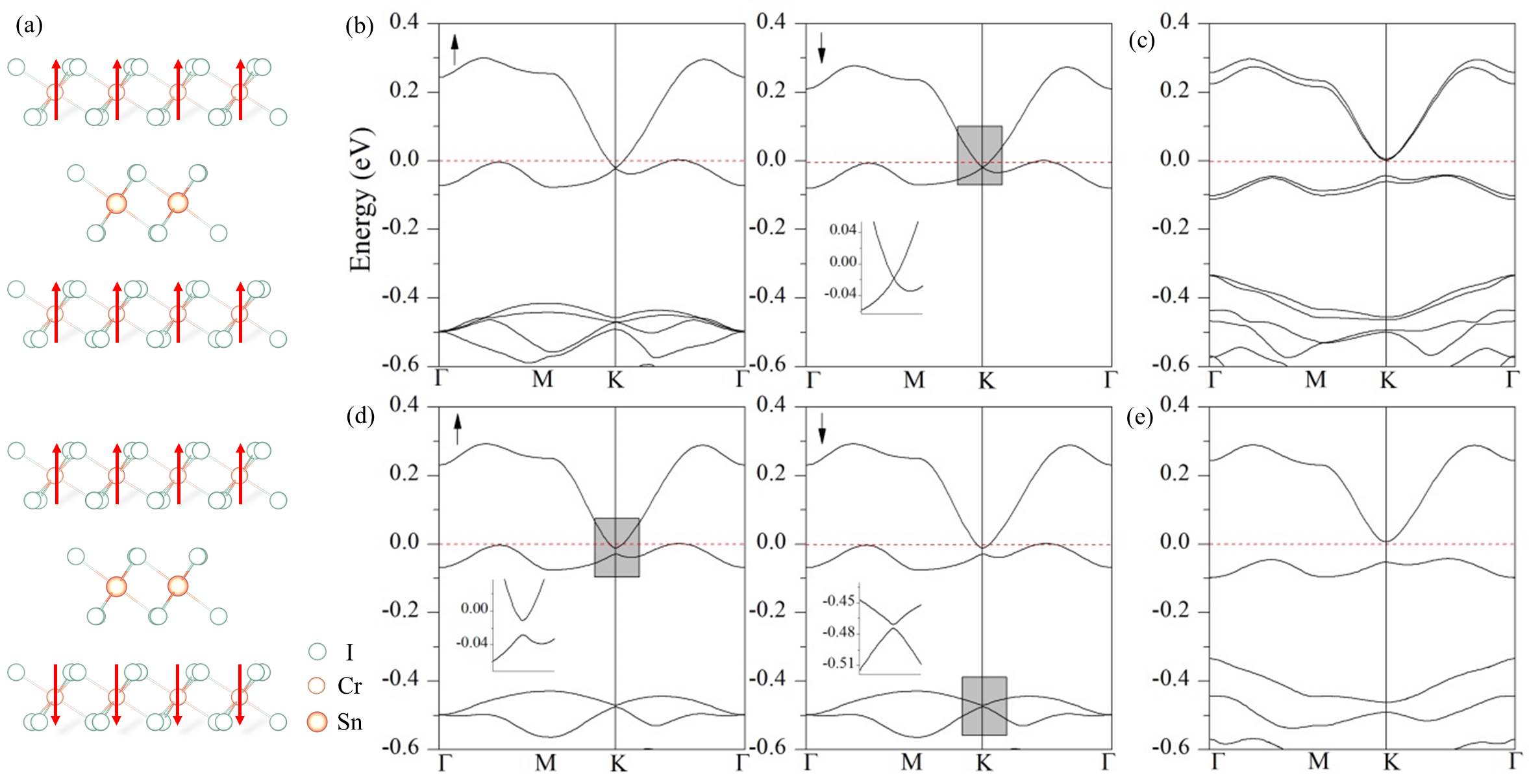}
	\centering
	\caption{(a) The illustration of ferromagnetic (top) and antiferromagnetic (bottom) alignment of the CrI$_3$/SnI$_3$/CrI$_3$ trilayer heterostructure. The ferromagnetic band structure without (b) and with (c) SOC. The antiferromagnetic band structure without (d) and with (e) SOC. In (b) and (d) the left and right panels correspond to spin-up and spin-down band structures, respectively. The insets show the band dispersion near the original Dirac point. }
	\label{fig:bs_FM_AFM}
\end{figure*}

The interplay between the magnetism in CrI$_3$ and the topological band structure in SnI$_3$ can be viewed from two perspectives --- the effect of magnetism on the band topology, and the effect of the large SOC on the magnetism. First let's look at the influence of magnetism on the band structure. Due to the weak van de Waals interaction, the two bands near $E_F$ are still from SnI$_3$. However, this weak interaction has some influence on the two bands, compared with the monolayer case (see Figure \ref{fig:band_structure}), that the bottom band raises higher along $K\Gamma$ than the Dirac point, leading to a compensated semimetal. The ferromagnetic spin-up band structure is different from the spin-down band structure due to the exchange potentials seen by spin-up and spin-down electrons are different\cite{Slater_mag}, manifesting in that several bands appear between $-0.4$ eV and $-0.6$ eV for spin-up electrons. The antiferromagnetic spin-up and spin-down band structures are the same, due to the equivalence of spin-up and spin-down electrons. The striking difference between the ferromagnetic and antiferromagnetic band structures is that, the Dirac point at $K$ is preserved under ferromagnetic alignment, while a gap (about $16$ meV) is opened under antiferromagnetic alignment, see Figure \ref{fig:bs_FM_AFM} (b) and (d). Furthermore, under antiferromagnetic alignment, only the double degeneracy at $K$ is split, while the double degeneracy at $\Gamma$ is preserved.

 This peculiar behavior can be understood as follows. The magnetic space groups of the heterostructure under ferromagnetic and antiferromagnetic alignments are $C_{3i}^1$ and $C_{3i}^{1}(C_3^1)$, respectively\cite{Cracknell,PhysRev.127.391}. In the former case the time-reversal symmetry $T$ is absent, while in the latter $T$ exists in combination with $I$, $S_6^{+}$ and $S_6^{-}$. By inspecting the character tables of the group of wave vector of $C_{3i}^1$, it is found that there are doubly degenerate representations at $\Gamma$ and $K$. Thus the Dirac point at $K$ is preserved under ferromagnetic alignment. However, the character tables of the group of wave vectors of space group $C_{3}^1$ show a tiny but important difference. At $\Gamma$, there are one non-degenerate and one doubly degenerate representation, which accounts for the preservation of double degeneration at $\Gamma$ under antiferromagnetic alignment. While at $K$, there are three representations, one is non-degenerate, the other two are also non-degenerate. This explains the splitting of Dirac point at $K$. However, the last two representations at $K$ have an extra ``degeneracy" with representations at $-K$, if $T$ is present. When $T$ is not present, as in our case, the examination of the magnetic space group of $C_{3i}^{1}(C_3^1)$ shows that this ``degeneracy" still exists, and we indeed find that the band structure along $\Gamma K$ is the same as that along $\Gamma (-K)$.

 The effect of large SOC on magnetism is revealed by the change of the magnetic moment of the unit cell. Under ferromagnetic alignment, the SnI$_3$ layer shows diamagnetism without SOC, and shows Van Vleck paramagnetism with SOC. This is confirmed by replacing the SnI$_3$ layer with BiI$_3$, under which case the magnetic moment of the unit cell is $12.0000 \mu_B$ (Note that there are 4 CrI$_3$ formula units per unit cell, this corresponds to $3.0000 \mu_B$ per Cr atom). This weak magnetism in SnI$_3$ monolayer is revealed by the band structure with SOC, as shown in Figure \ref{fig:bs_FM_AFM}(c), where the four bands near $E_F$ are not doubly degenerate. However, under antiferromagnetic alignment, no magnetism is introduced into the SnI$_3$ monolayer, and the four bands near $E_F$ are doubly degenerate, as in Figure \ref{fig:bs_FM_AFM} (e).  Weak paramagnetism is also seen in Bi$_2$Se$_3$\cite{Yu61}, and is supposed to open an gap for the edge state. The Dzyaloshinskii-Moriya interaction\cite{DZYALOSHINSKY1958241,PhysRev.120.91}, if exists in this system, is believed to be weak, as revealed by the tiny magnetic moment in $y$ direction under antiferromagnetic alignment.
\begin{table}
	\centering
	\caption{The magnetic moment per unit cell under ferromagnetic (FM) and antiferromagnetic (AFM) alignments for the heterostructure, in unit of $\mu_B$.}
	\label{tab:mag_mom}
	\begin{tabular}{c|cc}
		\hline
		alignment & without SOC & with SOC \\
		\hline
		FM & $11.9906$ & $(0.0001,0,12.0932)$ \\
		AFM & $0.0000$ & $(0,0.0002,0)$ \\
		\hline
		
	\end{tabular}
\end{table}
\subsection{Magnetic proximity effect induced topological phase transition}
Under the antiferromagnetic alignment of CrI$_3$ layers, the trilayer structure breaks both time-reversal and inversion symmetry, but preserves the symmetry of their combination $TP=i\sigma_x s_y K$. The perturbation satisfying these symmetry conditions is $H_{AFM}=\mu\sigma_z s_z$. So the effective Hamiltonian including both SOC and antiferromagnetism is
\begin{eqnarray}
  H_{eff}(\vec{k}) &=& v_F(k_x\sigma_x\tau_z+k_y\sigma_y)+\lambda \sigma_z\tau_z s_z+\mu\sigma_z s_z.
\end{eqnarray}
$H_{eff}$ can be block diagonalized in the eigenbasis of $s_z$, which means the spin-up and spin-down Hamiltonian can be separately studied. For each spin, the Hamiltonian is the same as the effective Hamiltonian of Haldane model, which includes the $T$-breaking term $\lambda \sigma_z\tau_z$ and the $P$-breaking term $\mu\sigma_z$. In Haldane model, the fermions are spinless, and the competition between the two terms leads to a topological phase transition between a trivial insulator and a Chern insulator. Here, since both spins are taken into account, the competition between the two terms leads to a topological phase transition between a trivial antiferromagnetic insulator and a TI. The phase transition occurs at $|\mu|=|\lambda|$. When $|\mu|>|\lambda|$, the system is a trivial insulator; when $|\mu|<|\lambda|$, it is a TI. The CrI$_3$/SnI$_3$/CrI$_3$ trilayer system belongs to the latter, since the antiferromagnetism is not large enough to surpass the SOC. We can expect that in other similarly layered systems, where the SOC is smaller or the antiferromagnetism is larger, the magnetic proximity effect will induce a topological phase transition. Note that since $s_z$ is conserved, although time-reversal symmetry is broken, the edge states do not hybridize and are still gapless.

\section{Conclusion}
\label{conclusion}
In conclusion, we have constructed a CrI$_3$/SnI$_3$/CrI$_3$ trilayer heterostructure model to study the magnetic proximity effect acting on a 2D TI. The SnI$_3$ monolayer is in BiI$_3$ structure. The band structure shows Dirac type dispersion near the Fermi energy when SOC is ignored, and a gap (about $0.11$ eV) opens after SOC is taken into account. The non-vanishing $Z_2$ invariant, together with the effective Hamiltonian same as Kane-Mele model, suggest that it is a $Z_2$ TI. We find that if the magnetic moments of Cr lie ferromagnetically between layers, the Dirac point is preserved; while if they lie antiferrromagnetically, a gap about $16$ meV is opened. This is explained by the fact that ferromagnetic and antiferromagnetic alignment lead to different magnetic space groups $C_{3i}^1$ and $C_{3i}^1(C_3^1)$. The gap opened by the magnetic proximity effect under the antiferromagnetic alignment competes with that opened by SOC, determining whether the system is a trivial antiferromagnetic insulator or a TI. Our work shows that magnetic proximity effect can tune a topological phase transition between a trivial insulator and a TI.

\ack
The authors would like to thank M. Z. Liu and Z. Yan for helpful discussions. D.Z. thanks the financial support by NSFC (No.$11574403$) and Guangzhou science and technology project (No.$201707020002$). Y.Z., L.W. and D.X.Y. are supported by the National Key R\&D Program of China 2017YFA0206203 and 2018YFA0306001, NSFC-11574404, the National Supercomputer Center in Guangzhou,and the Leading Talent Program of Guangdong Special Projects.

\section*{Appendix}
\label{sec:app}
The first-principles calculations were carried out by using the Vienna \emph{ab initio} simulation package (VASP)\cite{PhysRevB.54.11169}, with pseudopotentials constructed by projector augmented wave (PAW) method\cite{PhysRevB.50.17953}. The generalized gradient approximation (GGA) as suggested by Perdew, Burke and Ernzerhof (PBE)\cite{PhysRevLett.77.3865} was used to depict the exchange and correlation interaction. A cutoff energy of $700$ eV and a Monkhorst-Pack grid\cite{PhysRevB.13.5188} with $15\times15\times1$ were used to ensure convergence of 1 meV per unit cell for the total energy. The atoms are allowed to relax until the forces were smaller than 0.01 eV/\AA. In the phonon calculation the atoms were relaxed further until the forces were smaller than $10^{-8}$ eV/\AA, and Phonopy\cite{TOGO20151} was used to prepare the supercell and analyze the phonon spectrum. The zero-damping DFT-D3 method of Grimme\cite{doi:10.1063/1.3382344} was employed to describe the van der Waals interaction between the layers.

Here we give the calculation of $Z_2$ invariant. Due to the presence of inversion symmety, we can use Fu and Kane's method\cite{PhysRevB.76.045302} to calculate the $Z_2$ invariant by computing the parity eigenvalues of occupied levels at four time-reversal invariant momenta (TRIM). The results are listed in Table \ref{tab:parity}. According to their method, the $Z_2$ invariant $\nu$ is defined as
\begin{equation}
	(-1)^{\nu}=\prod_i\delta_i=\prod_i\prod_{m=1}^{N}\xi_{2m}(\Gamma_i)
\end{equation}
where $\Gamma_i$'s are the four TRIMs in the Brillouin zone,  $\xi_{2m}(\Gamma_i)=\pm1$ is the parity eigenvalue of the $2m$-th valence band at $\Gamma_i$. If $\nu=1$, the system is TI, while if $\nu=0$, the system is a trivial insulator. At the end of Table \ref{tab:parity} we give the sign of $\delta_i$. Since $\delta=1$ at the three $M$'s and $\delta=-1$ at $\Gamma$, we have $\nu=1$, confirming the SnI$_3$ monolayer is a TI.
\begin{table*}[ht]
\centering
\caption{The parity eigenvalues of the $2m$-th valence band at four TRIMs: $\Gamma(0,0,0)$,$M_1(1/2,0,0)$,$M_2(0,1/2,0)$ and $M_3(1/2,-1/2,0)$. The last second column is the parity of the bottom conduction band, and the last column is the value of $\delta$.}
\label{tab:parity}
\begin{tabular}{c|ccccccccccccccc}
	\hline
	2m & 2 & 4 & 6 & 8 & 10 & 12 & 14 & 16 & 18 & 20 & 22 & 24 & 26 & 28 & 30 \\
	\hline
	$M$ & $+1$ & $-1$ & $-1$ & $+1$ & $+1$ & $-1$ & $+1$ & $-1$ & $-1$ & $+1$ & $-1$ & $+1$ & $-1$ & $+1$ & $-1$ \\
	$\Gamma$ & $-1$ & $+1$ & $+1$ & $-1$ & $-1$ & $-1$ & $+1$ & $+1$ & $+1$ & $-1$ & $+1$ & $+1$ & $+1$ & $-1$ & $-1$ \\
	\hline	
	2m & 32 & 34 & 36 & 38 & 40 & 42 & 44 & 46 & 48 & 50 & 52 & 54 & 56 & 58 & 60 \\
	\hline
	$M$ & $+1$ & $+1$ & $-1$ & $-1$ & $+1$ & $-1$ & $+1$ & $-1$ & $+1$ & $-1$ & $+1$ & $-1$ & $+1$ & $-1$ & $+1$ \\
	$\Gamma$ & $-1$ & $-1$ & $+1$ & $+1$ & $+1$ & $-1$ & $+1$ & $-1$ & $+1$ & $-1$ & $+1$ & $-1$ & $+1$ & $+1$ & $-1$ \\
	\hline
	2m & 62 & 64 & 66 & 68 & 70 & 72 & $\delta_i$\\
	\hline
	$M$ & $-1$ & $+1$ & $-1$ & $+1$ & $-1$ & $+1$ & $+1$ \\
	$\Gamma$ & $-1$ & $+1$ & $-1$ & $-1$ & $+1$ & $-1$ & $-1$ \\
	\hline
\end{tabular}
\end{table*}

\section*{References}

\bibliography{SnI3}

\end{document}